\documentclass[runningheads]{llncs}          % LaTeX 2e

\usepackage{amsmath}

\usepackage{latexsym}
\usepackage{amsfonts}
\usepackage{amssymb}
\usepackage{mathtools}

\newcommand{\ar}[1]{\xrightarrow{#1}}

\newcommand{\dar}[1]{\xRightarrow{#1}}

\newcommand{\A}{\mathcal{A}}
\renewcommand{\P}{\mathcal{P}}
\newcommand{\Q}{\mathcal{Q}}
\newcommand{\G}{\mathcal{G}}
\newcommand{\R}{\mathcal{L}}

\newcommand{\TS}[1]{{\mathcal{TS}(#1)}}
\newcommand{\TT}[1]{{\mathcal{TT}(#1)}}

\newcommand{\ST}[1]{{\mathcal{ST}(#1)}}

\newcommand{\para}{\|}

\usepackage[zed]{csp}

\begin{document}

\title{Revisiting Timed Specification Theories: \\A Linear-Time Perspective}

\author{Chris Chilton \and Marta Kwiatkowska \and Xu Wang}

\institute{Department of Computer Science, University of Oxford, UK}

\maketitle

\vspace{-4mm}

\begin{abstract}
We consider the setting of component-based design for real-time systems with critical timing constraints. Based on our earlier work, we propose a compositional specification theory for timed automata with I/O distinction, which supports substitutive refinement. Our theory provides the operations of parallel composition for composing components at run-time, logical conjunction/disjunction for independent development, and quotient for incremental synthesis. The key novelty of our timed theory lies in a weakest congruence preserving safety as well as bounded liveness properties. We show that the congruence can be characterised by two linear-time semantics, \emph{timed-traces} and \emph{timed-strategies}, the latter of which is derived from a game-based interpretation of timed interaction.

\end{abstract}

\section{Introduction}
\label{intro}

Component-based design methodologies can be encapsulated in the form of compositional specification theories, which allow the mixing of specifications and implementations, admit substitutive refinement to facilitate reuse, and provide a rich collection of operators. %, including parallel composition to reason about component behaviour at run-time, conjunction to enable independent development constrained by several specifications, and quotienting for incremental synthesis.
%Component-based design is a popular methodology used in the development of real-time systems with critical timing constraints. It provides us with an excellent opportunity to exploit compositional reasoning on timed components, e.g. assume/guarantee and interface/compatibility techniques, which are widely believed to hold the promise of scalable analysis. %Previously
Previously \cite{ESOP}, we developed a linear-time specification theory for reasoning about untimed components that interact by synchronisation of input and output (I/O) actions, inspired by interface automata~\cite{henzinger-ia}. Models can be specified operationally by means of transition systems augmented by an inconsistency predicate on states, or declaratively using traces. The theory admits non-determinism, a refinement preorder based on traces, and the operations of parallel composition, conjunction and quotient. The refinement is strictly weaker than alternating simulation and is actually the weakest pre-congruence preserving inconsistent states. This implies that our refinement is \emph{substitutive}, meaning component $A$ refines component $B$ iff $A$ can replace $B$ in any environmental context without introducing additional errors.

In this paper we target component-based development for real-time systems with critical timing constraints. We formulate a timed extension of the linear-time specification theory of \cite{ESOP}, by allowing for both operational descriptions of components, as well as declarative specifications based on traces. Our operational models are based on a variant of timed automata with I/O distinction (although we do not insist on input-enabledness, cf~\cite{Kaynar}), augmented by two special states: $\bot$ for safety and bounded-liveness errors, and $\top$ for timestop. Trace-based declarative specifications are shown to be a suitable semantic domain for the operational models. In addition to timed-trace semantics, we present timed-strategy semantics, which coincides with the former but relates our work closer to the timed-game frameworks used by~\cite{larsen-timedio} and~\cite{henzinger-timedia}. %\marginpar{Chris: two levels require clarification} Furthermore, unlike untimed cases, in timed specification theory we have two different levels of timed systems: one is the plain timed systems while the other is the timed systems that are empowered with the capability of \emph{timestop} (i.e. to stop time).
The \emph{substitutive refinement} of our framework %based on traces
%Our specification theory is mostly based on the level without timestop. We show the definition of specification refinement actually builds on that of the refinement on the level with timestop. It is a coarsening of the latter since the loss of timestop reduces the distinguishing power of the context as well. Both refinements are pre-congruences on their respective levels.
%As for the untimed case, the refinement
gives rise to the weakest congruence preserving $\bot$, and is shown to coincide across all our formalisms. %t states, which in the timed case models both safety errors (aka immediate errors in~\cite{henzinger-timedia}) and bounded-liveness errors (aka time errors in~\cite{henzinger-timedia}).

Amongst notable works in the literature, we briefly mention a theory of timed interfaces~\cite{henzinger-timedia} and a theory of timed specifications~\cite{larsen-timedio}. Timed interface theory contributes a framework based on timed games to formalise notions such as interfaces and compatibility, and also provides a parallel composition operator. However, the work cannot be considered a specification theory as it does not deal with the notion of refinement for component substitution or the operations of conjunction, disjunction and quotient. In this respect,~\cite{larsen-timedio} provides a complete theory; however, the refinement is a timed version of the alternating simulation originally defined for interface automata~\cite{henzinger-ia}. Consequently, it is too strong for determining when a component can be safely substituted with another (cf the example in Figure~\ref{fig:strategy-equivalence-strong}).

\paragraph{Outline.}%The organisation of the paper is as follows.
In Section~\ref{frame} we introduce timed I/O automata, their semantic mapping to timed I/O transition systems, and supply the operational definitions for the operations of parallel composition, conjunction, disjunction and quotient.
% Compared to classical definitions, the novelty of our approach lies in the new notion of co-invariants and the introduction of inconsistent states in TIOTSs.
In Section~\ref{sec:game} we use the timed-game framework to introduce timed-strategy semantics, which we relate to the operational framework. Similarly in Section~\ref{sec:ts}, we present timed-trace semantics and relate these to the operational definitions. %The semantics is developed first on the level with timestop, and then refined by adding realisability conditions to remove explict and implict timestops for the other level. \marginpar{Chris: do we even deal with realisability now?}Then in Section\label{sec:ts} we give the second semantics of \emph{time-traces}.
%We show that trace semantics corresponds to the timed-strategies semantics.
Section~\ref{sec:comp} discusses related work, and finally Section~\ref{sec:concl} concludes.

\section{Formal Framework}
\label{frame}

In this section we introduce timed I/O automata, timed I/O transition systems and a semantic mapping from the former to the latter. Timed I/O automata are compact representations of timed I/O transition systems. Our theory will be developed using timed I/O transition systems, which are endowed with a richer repertoire of semantic machinery.

\subsection{Timed I/O Automata}\label{sec:time}

\paragraph{Clock constraints.} Given a set $X$ of real-valued clock variables, a \emph{clock constraint} over $X$, $cc: CC(X)$, is a boolean combination of atomic constraints of the form $x \bowtie d$ and $x - y \bowtie d$ where $x,y \in X$, $\bowtie \in \{\leq,<, =, >,\geq\}$, and $d \in \mathbb{N}$.

A \emph{clock valuation} over $X$ is a map $t$ that assigns to each clock variable $x$ in $X$ a real value from $\mathbb{R}^{\geq 0}$. We say $t$ satisfies $cc$, written $t \in cc$, if $cc$ evaluates to true under valuation $t$. $t + d$ denotes the valuation derived from $t$ by increasing the assigned value on each clock variable by $d \in \mathbb{R}^{\geq 0}$ time units. $t[rs \mapsto 0]$ denotes the valuation obtained from $t$ by resetting the clock variables in $rs$ to $0$. Sometimes we use $0$ for the clock valuation that maps all clock variables to $0$. %We can use $(t,t')$, $(t,t']$, $[t,t')$, and $[t,t']$ to denote intervals of clock valuations if $t' = t + d$ for some $d \in \mathbb{R}^{\geq 0}$. Note that we use the convention that $(t,t]$ and $[t,t)$ both denote the singleton $\{t\}$.

\begin{definition}
A \emph{timed I/O automaton} (TIOA) is a tuple $(C, I, O, L, l^0, AT,$ $ Inv, coInv)$, where:

\begin{itemize}
\item $C \subseteq X$ is a finite set of clock variables

\item $A$ ($= I \cup O$) is a finite alphabet, where $I$ and $O$ are disjoint sets of input actions and output actions respectively

\item $L$ is a finite set of \emph{locations}

\item $l^0 \in L$ is the \emph{initial location}

\item $AT \subseteq L \cross CC(C) \cross A \cross 2^{C} \cross L$ is a set of \emph{action transitions}

\item $Inv : L \fun CC(C)$ and $coInv : L \fun CC(C)$ assign \emph{invariants} and \emph{co-invariants} to states, each of which is a downward-closed clock constraint.
\end{itemize}
\end{definition}

We use $l,l',l_i$ to range over $L$ and use $l \ar{g,a,rs} l'$ as a shorthand for $(l, g, a, rs, $ $l') \in AT$. $g: CC(C)$ is the enabling guard of the transition, $a \in A$ the action, and $rs$ the subset of clock variables to be reset.

%The semantics of our TIOAs are largely those of finite word safety timed automata with point-wise and weakly monotonic semantics.
Our TIOAs are similar to existing variants of timed automata with input/output distinction, except for the introduction of co-invariants and non-insistence on input-enabledness. While invariants specify the bounds beyond which time may not progress, co-invariants specify the bounds beyond which the system will \emph{time-out} and enter error states. Our TIOAs can be used to describe both the assumptions made by the component on the inputs, together with the guarantees provided by the component on the outputs. Such assumptions and guarantees can be time constrained: guards on output transitions express safety timing guarantees, while guards on input transitions express safety timing assumptions; invariants (urgency) express liveness timing guarantees on outputs while co-invariants (time-out) express liveness timing assumptions on inputs.

When components interact together, we check whether the guarantees they provide meet the assumptions they make on each other. If not, there are two types of errors:

\begin{itemize}
\item An input arrives in a state and at a time when it is not expected (i.e. not satisfying the guards on the input transitions). This is a \emph{safety error}. % (aka immediate error in~\cite{henzinger-timedia})
\item An input does not arrive in a state within a time bound (specified by a co-invariant) as expected. This is a \emph{bounded-liveness error}. % (aka time error in~\cite{henzinger-timedia})
\end{itemize}

\begin{figure}[t]
\begin{center}
\includegraphics[width=\textwidth]{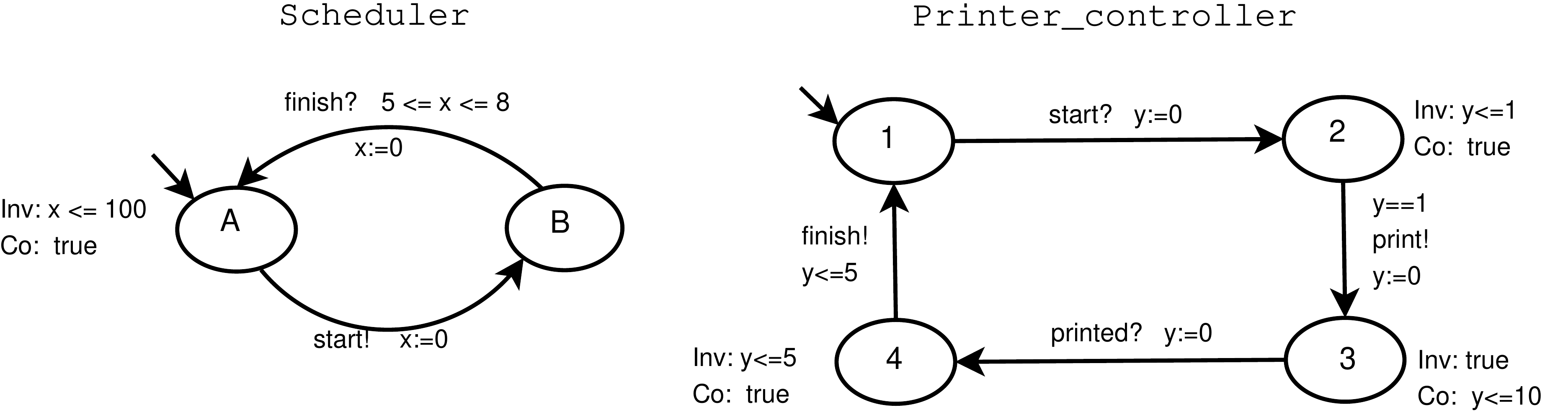}
\end{center}
\vspace{-4mm}
\caption{Job scheduler and printer controller.}
\vspace{-5mm}
\label{fig:automata}
\end{figure}

\paragraph{Example.} Figure~\ref{fig:automata} depicts TIOAs representing a job scheduler together with a printer controller. The invariant at location $A$ of the scheduler forces a bounded-liveness guarantee on outputs in that location. As time must be allowed to progress beyond $t=100$, the $start$ action must be fired within the range $0\leq t \leq 100$. After $start$ has been fired, the clock $x$ is reset to $0$ and the scheduler waits (possibly indefinitely) for the job to $finish$. If the job does finish, the scheduler is only willing for this to take place between $5\leq t \leq 8$ after the job started (safety assumption), otherwise an unexpected input error will be thrown.

The controller waits for the job to $start$, after which it will wait exactly $1$ time unit before issuing $print$ (forced by the invariant $y\leq 1$ on state $2$ and the guard $y=1$). The controller now requires the printer to indicate the job is $printed$ within $10$ time units of being sent to the printer, otherwise a time-out error on inputs will occur (co-invariant $y\leq 10$ in state $3$ as liveness assumption). After the job has finished printing, the controller must indicate to the scheduler that the job has $finish$ed within $5$ time units.

\subsection{Timed Actions and Words}

In this section we introduce some notation relating to timed actions and timed words that will be of use to us in later sections.

\vspace{-2mm}

\paragraph{Timed actions.} For a set of input actions $I$ and a set of output actions $O$, define $tA= I \uplus O \uplus \mathbb{R}^{>0}$ to be the set of \emph{timed actions}, $tI= I  \uplus \mathbb{R}^{>0}$ to be the set of \emph{timed inputs}, and $tO= O \uplus \mathbb{R}^{>0}$ to be the set of \emph{timed outputs}. We use symbols like $\alpha$, $\beta$, etc. to range over $tA$.

\vspace{-2mm}

\paragraph{Timed words.} A \emph{timed word} (ranged over by $w, w', w_i$ etc.) is a finite mixed sequence of positive real numbers ($\mathbb{R}^{> 0}$) and visible actions such that \emph{no two numbers are adjacent to one another}. For instance, $\langle 0.33, a, 1.41, b, c, 3.1415 \rangle$ is a timed word denoting the observation that action $a$ occurs at $0.33$ time units, then another $1.41$ time units lapse before the simultaneous occurrence of $b$ and $c$, which is followed by $3.1415$ time units of no event occurrence. The empty word is denoted by $\epsilon$. %More generally, zero can be freely inserted into a timed word as long as the non-adjacency of numbers is not violated. For example, $\langle a, 1.41, b, c \rangle$ is equivalent to $\langle 0, a, 1.41, b, 0, c, 0\rangle$.

\vspace{-2mm}

\paragraph{Operations on timed words.} We use $last(w)$ to denote the last element in the sequence $w$, and $l(w)$ to indicate the length, which is obtained as the sum of all the reals in $w$. Concatenation of timed words $w$ and $w'$ is obtained by appending $w'$ onto the end of $w$ and coalescing adjacent reals (summing them). For instance, $\langle a, 1.41 \rangle$ $\cat \langle 0.33, b, 3.1415 \rangle$ = $\langle a, (1.41 + 0.33), b, 3.1415 \rangle$ = $\langle a, 1.74, b, 3.1415 \rangle$. Prefix/extension are defined as usual by concatenation, and we use $\leq$ for the prefix partial order. We write $w \upharpoonright tA_0$ for the projection of $w$ onto timed alphabet $tA_0$, which is defined by removing from $w$ all actions not inside $tA_0$ and coalescing adjacent reals. %Furthermore we say $w'$ is a \emph{strong prefix} of $w$ if $w= w_0 \cat \langle last(w) \rangle$ and $w'$ is a prefix of $w_0$.

\subsection{Semantics as Timed I/O Transition Systems}
\label{sec:semantics}

The semantics of TIOAs are given as timed I/O transition systems, which are a special class of infinite labelled transition systems.

\begin{definition}
A \emph{timed I/O transition system} (TIOTS) is a tuple $\P= \langle I, O, S,$ $ s^0, \rightarrow \rangle$, where: $I$ and $O$ are the input and output actions respectively, $S$ is a set of states, $s^0$ is the designated initial state, and $\rightarrow \subseteq S \cross I \uplus O \uplus \mathbb{R}^{>0} \cross S$ is the action and time-labelled transition system.
\end{definition}

The states of the TIOTS for a TIOA capture the configurations of the automaton, i.e. its location and clock valuation. Therefore, each state of the TIOTS is a pair drawn from $L\times \mathbb{R}^{C}$, which we refer to as the set of \emph{plain states}, denoted $P$. In addition, we introduce two special states $\bot$ and $\top$, which are required for the semantic mapping of disabled inputs/outputs, invariants and co-invariants.

$\bot$ is called the \emph{inconsistent state}, representing safety and bounded-liveness errors. $\top$ is the so-called \emph{timestop state}, representing the \emph{magic moment} from which no error can occur.\footnote{For instance, a location with $true$ as co-invariant and $false$ as invariant is mapped to $\top$, while a location with $true$ as invariant and $false$ as co-invariant is mapped to $\bot$. A location with $false$ for both invariant and co-invariant is mapped to $\top$ since invariants have priority over co-invariants according to our semantics; whereas a location with $x \leq 0$ as invariant and $true$ as co-invariant is mapped to a plain state.}

%The trivial TIOTS with $\top$ (resp. $\bot$) as the initial state is called the $\top$-TIOTS (resp. $\bot$-TIOTS).

An intuitive way to understand $\top$ and $\bot$ is from an input/output game perspective. The component controls output and delay while the environment controls input. $\bot$ is the losing state for the environment. So a disabled input at a state $p$ is equated to an input transition from $p$ to $\bot$.  $\top$ is the losing state for the component. So a disabled output/delay at $p$ is equated to an output/delay transition from $p$ to $\top$. Thus we can have two semantics-preserving transformations on TIOTSs.

The \emph{$\bot$-completion} of a TIOTS $\P$, denoted $\P^{\bot}$, adds an $a$-labelled transition from $p$ to $\bot$ for every $p \in P_{\P}$ and $a \in I$ s.t. $a$ is not enabled at $p$. $\bot$-completion will make a TIOTS \emph{input-receptive}, i.e. input-enabled at all states. The \emph{$\top$-completion} of a TIOTS $\P$, denoted $\P^{\top}$, adds an $\alpha$-labelled transition from $p$ to $\top$ for every $p \in P_{\P}$ and $\alpha \in tO$ s.t. $\alpha$ is not enabled at $p$. %$\top$-completion will make a TIOTS \emph{input-receptive}, i.e. input-enabled at all states.

%start with $\P$, then for each $p \in P_{\P}$ and $a \in I$, if $a$ is not enabled at $p$, we add . Unlike the work in~\cite{aalborg}, we do not assume TIOTSs are \emph{receptive} or input-enabled at all states.

%We use $\P^{\top}$ to denote the \emph{$\top$-completion} of $\P$, which transforms TIOTS $\P$ in the following manner: for all $p \in P_{\P}$ and $\alpha \in tO$, if $\alpha$ is not enabled at $p$, we add an $\alpha$-labelled transiton from $p$ to $\top$. Like $\bot$-completion, $\top$-completion leaves the semantics of a TIOTS unchanged.

%}, i.e. the late arrival of expected inputs. (The latter can be implemented by the time-out of co-invariants.)

Furthermore, for technical convenience (e.g. ease of defining time additivity), the definition of TIOTSs requires that 1) $\top$ is a \emph{quiescent state}, i.e. a state in which the set of outgoing transitions are all self-loops, one for each $d \in \mathbb{R}^{>0}$, and 2) $\bot$ is a \emph{chaotic state}, i.e. a state in which the set of outgoing transitions are all self-loops, one for each $\alpha \in tA$. The set of all possible states is denoted $S = P \uplus \{\bot,\top\}$. We use $p,p',p_i$ to range over $P$ while $s,s',s_i$ range over $S$.

The transition relation $\rightarrow$ of the TIOTS is derived from the execution semantics of the TIOA.

\begin{definition}
Let $\P$ be a TIOA. The semantic mapping of $\P$ is a TIOTS $\langle I, O, S, s^0, \rightarrow \rangle$, where:

\begin{itemize}
\item $S = (L \times \mathbb{R}^C) \uplus \{\bot, \top\}$
\item $s^0 = \top$ providing $0 \notin Inv(l^0)$, $s^0 = \bot$ providing $0 \in  Inv(l^0) \wedge \neg coInv(l^0)$ and $s^0 = (l^0, 0)$ providing $0 \in  Inv(l^0) \wedge coInv(l^0)$,
\item $\rightarrow$ is the smallest relation satisfying:
\begin{enumerate}
\item If $l \ar{g,a,rs} l'$, $t'= t [rs \mapsto 0]$, $t \in Inv(l) \wedge coInv(l) \wedge g$, then:
\begin{enumerate}
\item \emph{plain action:} $(l, t) \ar{a} (l', t')$ providing $t' \in Inv(l') \wedge coInv(l')$
\item \emph{error action:} $(l, t) \ar{a} \bot$ providing $t' \in Inv(l') \wedge \neg coInv(l')$
\item \emph{magic action:} $(l, t) \ar{a} \top$ providing $t' \in \neg  Inv(l')$.
\end{enumerate}
\item \emph{plain delay:} $(l, t) \ar{d} (l, t+d)$ if $t,t+d \in Inv(l) \wedge coInv(l)$
\item \emph{time-out delay:} $(l, t) \ar{d} \bot$ if $t \in Inv(l) \wedge coInv(l)$, $t+d \notin coInv(l)$ and $\exists 0 < \delta \leq d: t+\delta \in Inv(l) \wedge \neg coInv(l)$.
\end{enumerate}
\end{itemize}
\end{definition}

%since time will stop at the very moment such an entry is made. So $\top$ captures only a subset of the possible timestops in a system, which we call explicit timestop.

Note that our semantics tries to minimise the use of transitions leading to $\top/\bot$ states. Thus there are no delay transitions leading to $\top$. This creates implicit timestops, which we capture using the concept of \emph{semi-timestop} (i.e. semi-$\top$). We say a plain state $p$ is a \emph{semi-$\top$} iff 1) all output transitions enabled in $p$ or any of its time-passing successors lead to the $\top$ state, and 2) there exists $d \in \mathbb{R}^{>0}$ s.t. $p \ar{d} \top$ or $d$ is not enabled in $p$. Thus a semi-$\top$ is a state in which it is impossible for the component to avoid the timestop without suitable inputs from the environment.

%Since timestop cannot be implemented by realistic systems, we say a TIOTS is \emph{realisable} iff it is free of timestop and semi-timestop, i.e. no semi-$\top$ or $\top$ state is reachable.

\paragraph{TIOTS terminology.} A TIOTS is \emph{time additive} providing
%The TIOTSs satisfies \emph{time additivity}, meaning
$p \ar{d_1+d_2} s'$ iff $p \ar{d_1} s$ and $s \ar{d_2} s'$ for some $s$. In the sequel of this paper we only consider TIOTSs that are time-additive.

We say a TIOTS is \emph{deterministic} iff there is no ambiguous transition in the TIOTS, i.e. $s \ar{\alpha} s' \wedge s \ar{\alpha} s''$ implies $s'=s''$.  %Other than the trivial non-realisable TIOTS (i.e. the one with $\top$ as the initial state) , our specification theory deal only with realisable TIOTSs.

Given a TIOTS $\P$, a timed word can be derived from a finite execution of $\P$ by extracting the labels in each transition and coalescing adjacent reals. The timed words derived from such executions are called \emph{traces} of $\P$. We use $tt, tt', tt_i$ to range over the set of traces and use $s^0 \dar{tt} s$ to denote a finite execution that produces trace $tt$ and leads to $s$.%The language of a specification is its set of timed traces, which is obviously prefix-closed and feasible. %For the magic specification, it is defined to have an empty set of traces.

\subsection{Operational Specification Theory}
\label{sec:optheory}

In this section we develop a compositional specification theory for TIOTSs based on the operations of parallel composition $\parallel$, conjunction $\wedge$, disjunction $\vee$ and quotient $\%$. The operators are defined via transition rules that are a variant on synchronised product.

Parallel composition yields a TIOTS that represents the combined effect of its operands interacting with one another. The remaining operations must be explained with respect to a refinement relation, which corresponds to safe-substitutivity in our theory. A TIOTS is a refinement of another if it will work in any environment that the original worked in without introducing safety or bounded-liveness errors. Conjunction yields the coarsest TIOTS that is a refinement of its operands, while disjunction yields the finest TIOTS that is refined by both of its operands. The operators are thus equivalent to the join and meet operations on TIOTSs\footnote{As we write $A\sqsubseteq B$ to mean $A$ is refined by $B$, our operators $\wedge$ and $\vee$ are reversed in comparison to the standard symbols for meet and join.}. Quotient is the adjoint of parallel composition, meaning that $\P_0 \% \P_1$ is the coarsest TIOTS such that $(\P_0 \% \P_1) \para \P_1$ is a refinement of $\P_0$.

Let $\P_i= \langle I_i, O_i, S_i, s_i^0, \rightarrow_i \rangle$ for $i \in \{0,1\}$ be two TIOTSs that are both $\bot$ and $\top$-completed, satisfying (wlog) $S_0\cap S_1 = \{\bot,\top\}$. The composition of $\P_0$ and $\P_1$ under the operation $\otimes\in \{\parallel, \wedge, \vee, \%\}$, written $\P_0 \otimes \P_1$, is only defined when certain \emph{composability} restrictions are imposed on the alphabets of the TIOTSs. $\P_0\parallel \P_1$ is only defined when the output sets of $\P_0$ and $\P_1$ are disjoint, because an output should be controlled by at most one component. Conjunction and disjunction are defined only when the TIOTSs have \emph{identical alphabets} (i.e. $O_0 = O_1$ and $I_0 = I_1$). This restriction can be relaxed at the expense of more cumbersome notation, which is why we focus on the simpler case in this paper. For the quotient, we require that the alphabet of $\P_0$ \emph{dominates} that of $\P_1$ (i.e. $A_1 \subseteq A_0$ and $O_1 \subseteq O_0$), in addition to $\P_1$ being a deterministic TIOTS. As quotient is a synthesis operator, it is difficult to give a definition using just \emph{state-local} transition rules, since quotient needs global information of the transition systems. This is why we insist on $\P_1$ being deterministic\footnote{Technically speaking, the problem lies in that state quotient operator is right-distributive but not left-distributive over state disjunction (cf Table~\ref{table:composition}).}.

\begin{table}[t]
\caption{State representations under composition operators.}

\begin{minipage}[b]{0.22\linewidth}
\centering
\begin{tabular}{l | l c l}
$\parallel$ &  $\top$ & $p_0$ & $\bot$ \\
\hline
$\top$ &  $\top$ & $\top$ & $\top$ \\
$p_1$ &  $\top$ & $p_0 \!\!\cross\!\! p_1$ & $\bot$ \\
$\bot$ &  $\top$ & $\bot$ & $\bot$
\end{tabular}
\end{minipage}
\hspace{0.2cm}
\begin{minipage}[b]{0.22\linewidth}
\centering
\begin{tabular}{l | l c l}
$\wedge$ &  $\top$ & $p_0$ & $\bot$ \\
\hline
$\top$ &  $\top$ & $\top$ & $\top$ \\
$p_1$ &  $\top$ & $p_0 \!\!\cross\!\! p_1$ & $p_1$ \\
$\bot$ &  $\top$ & $p_0$ & $\bot$
\end{tabular}
\end{minipage}
\hspace{0.2cm}
\begin{minipage}[b]{0.22\linewidth}
\centering
\begin{tabular}{l | l c l}
$\vee$ &  $\top$ & $p_0$ & $\bot$ \\
\hline
$\top$ &  $\top$ & $p_0$ & $\bot$ \\
$p_1$ &  $p_1$ & $p_0 \!\!\cross\!\! p_1$ & $\bot$ \\
$\bot$ &  $\bot$ & $\bot$ & $\bot$
\end{tabular}
\end{minipage}
\hspace{0.2cm}
\begin{minipage}[b]{0.22\linewidth}
\centering
\begin{tabular}{l | l c l}
$\%$ &  $\top$ & $p_0$ & $\bot$ \\
\hline
$\top$ &  $\bot$ & $\bot$ & $\bot$ \\
$p_1$ &  $\top$ & $p_0 \!\!\cross\!\! p_1$ & $\bot$ \\
$\bot$ &  $\top$ & $\top$ & $\bot$
\end{tabular}
\end{minipage}
\vspace{-5mm}
\label{table:composition}
\end{table}

\begin{definition}
Let $\P_0$ and $\P_1$ be TIOTSs composable under $\otimes \in \{\parallel, \wedge, \vee, \%\}$. Then $\P_0\otimes \P_1 = \langle I, O, S, s^0, \rightarrow\rangle$ is the TIOTS where:

\begin{itemize}
\item If $\otimes = \parallel$, then $I = (I_0\cup I_1) \setminus O$ and $O=O_0\cup O_1$
\item If $\otimes \in \{ \wedge,\vee\}$, then $I = I_0 = I_1$ and $O=O_0=O_1$
\item If $\otimes =\%$, then $I = I_0 \cup O_1$ and $O=O_0\setminus O_1$
\item $S = P_0 \times P_1 \uplus P_0 \uplus P_1 \uplus \{\top, \bot\}$
\item $s^0 = s^0_0 \otimes s^0_1$
\item $\rightarrow$ is the smallest relation containing $\rightarrow_0 \cup \rightarrow_1$, and satisfying the rules:
\begin{center}
\Large
\noindent
${p_0 \ar{\alpha}_{0} s_0'} \ \ {p_1 \ar{\alpha}_{1} s_1'} \over {p_0 \otimes p_1 \ar{\alpha} s_0' \otimes s_1'}$  \ \
\vspace{2mm} ${p_0  \ar{a}_0 s_0'} \ \  a \notin A_1  \over {p_0 \otimes p_1} \ar{a} {s_0' \otimes p_1}$ \ \
\vspace{2mm} ${p_1  \ar{a}_0 s_1'} \ \  a \notin A_0  \over {p_0 \otimes p_1} \ar{a} {p_0 \otimes s_1'}$ \ \
\normalsize
\end{center}
\vspace{-5mm}
\end{itemize}

\noindent We adopt the notation of $s_0 \otimes s_1$ for states, where the associated interpretation is supplied in Table~\ref{table:composition}. Furthermore, given two plain states $p_i=(l_i, t_i)$ for $i \in \{0,1\}$, we define $p_0 \times p_1 = ((l_0,l_1),t_0 \uplus t_1)$.
\end{definition}
%, as $\top$ can only be reached by an output or time-delay, meaning the corresponding transition in the other component will never be executed.
%If neither of the states are magic, then error takes precedence.

Table~\ref{table:composition} tells us how states should be combined under the
composition operators. From the environment's point of view, $\top$ refines
plain states, which in turn refines $\bot$. For parallel, a state is magic if
one component state is magic, and a state is error if one component is error
while the other is not magic. For conjunction, encountering error in one
component implies the component can be discarded and the rest of the composition behaves like the other component. The conjunction
table follows the intuition of the join operation on the refinement preorder.
Similarly for disjunction. Quotient is the adjoint of parallel composition. If the second component state does not refine the first, the quotient
will try to rescue the refinement by producing $\top$ (so that its
composition with the second will refine the first). If the second component state does refine the first, the quotient
will produce the least refined value so that its composition with the second will not break the refinement.

\

An \emph{environment} for a TIOTS $\P$ is any TIOTS $\Q$ such that the alphabet of $\Q$ is \emph{complementary} to that of $\P$, meaning $I_\P=O_\Q$ and $O_\P=I_\Q$. Refinement in our framework corresponds to contextual substitutability, in which the context is an arbitrary environment.

\begin{definition}
\label{defn:op-refine}
Let $\P_{imp}$ and $\P_{spec}$ be TIOTSs with identical alphabets. $\P_{imp}$ \emph{refines} $\P_{spec}$, denoted $\P_{spec} \sqsubseteq \P_{imp}$, iff for all environments $\Q$, $\P_{spec} \parallel \Q$ is $\bot$-free implies $\P_{imp} \parallel \Q$ is $\bot$-free. We say $\P_{imp}$ and $\P_{spec}$ are \emph{substitutively equivalent}, i.e. $\P_{spec} \simeq \P_{imp}$, iff $\P_{imp} \sqsubseteq \P_{spec}$ and $\P_{spec} \sqsubseteq \P_{imp}$.
\end{definition}

It is obvious that $\simeq$ induces an equivalance on TIOTSs and no equivalence that preserves the $\bot$ state can be weaker than $\simeq$. In the sequel we will give two concrete characterisations of $\simeq$ and show that $\simeq$ is also a congruence w.r.t. the parallel composition, conjunction, disjunction and quotient operators.

The operational definition of quotient requires that $\P_1$ is determinised, which can be accomplished by
a modified subset construction procedure on $(\P_1^{\bot})^{\top}$. If the current state subset $S_0$ contains $\bot$, it reduces $S_0$ to $\bot$; if $\bot \notin S_0 \neq \{\top\}$, it reduces $S_0$ by removing any potential $\top$ in $S_0$. As expected, the determinisation of $\P$, denoted $\P^D$, is substitutively equivalent to $\P$.

\begin{proposition}
Any TIOTS is substitutively equivalent to a deterministic TIOTS.
\end{proposition}

Equipped with determinisation, quotient is a fully defined operator on any pair of TIOTSs. Furthermore, we can give an alternative (although substitutively equivalent) formulation of quotient as the derived operator $(\P_0^{\neg}  \parallel \P_1)^{\neg}$, where $\neg$ is a mirroring operation that first determinises its argument, then interchanges the input and output sets, as well as the $\top$ and $\bot$ states.

\begin{figure}[t]
\begin{center}
\includegraphics[width=0.7\textwidth]{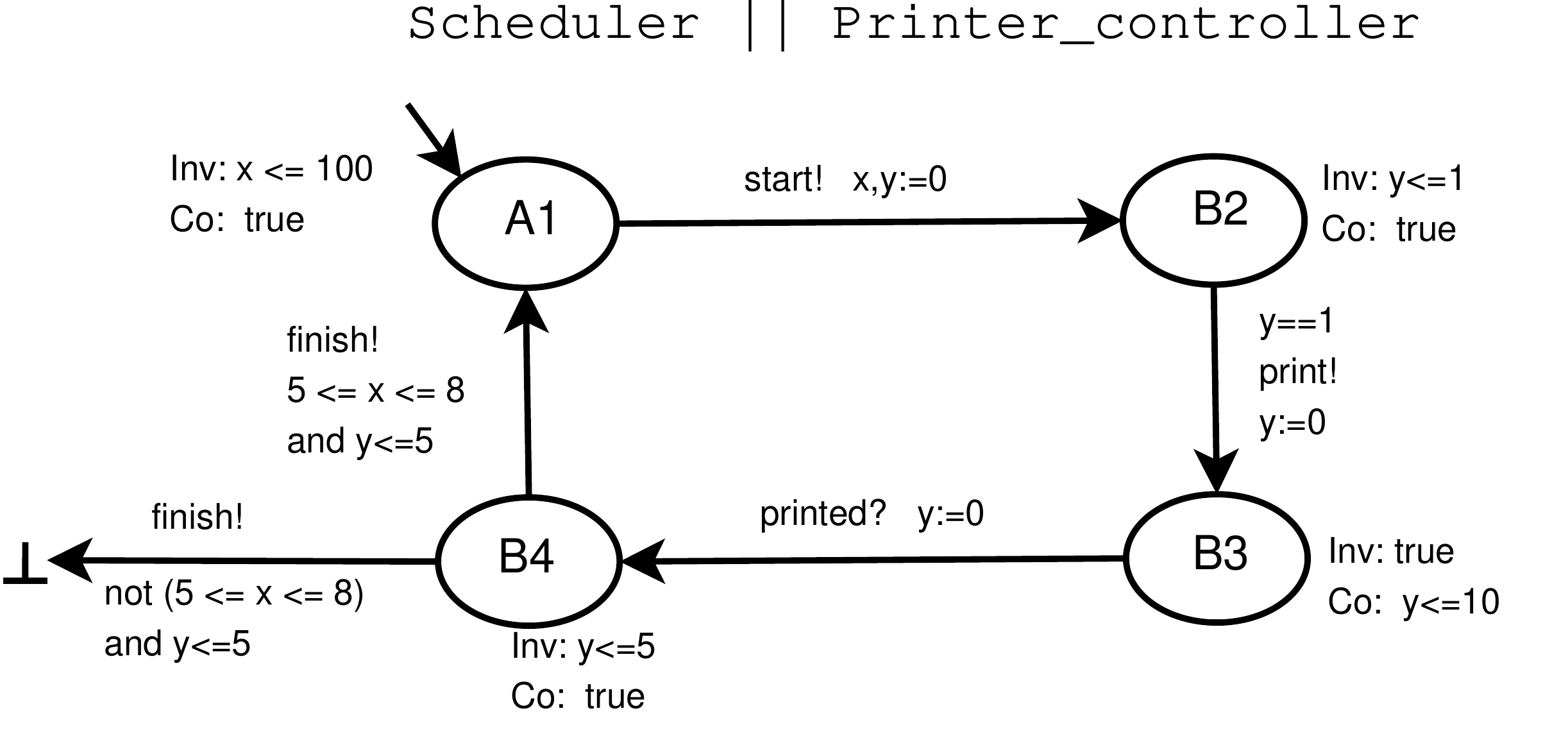}
\end{center}
\vspace{-5mm}
\caption{Parallel composition of the job scheduler and printer controller.}
\vspace{-5mm}
\label{fig:product}
\end{figure}

\paragraph{Example.} Figure~\ref{fig:product} shows the parallel composition of the job scheduler with the printer controller. In the transition from $B4$ to $A1$, the guard combines the effects of the constraints on the clocks $x$ and $y$. As $finish$ is an output of the controller, it can be fired at a time when the scheduler is not expecting it, meaning that a safety error will occur. This is indicated by the transition to $\bot$ when the guard constraint $5\leq x \leq 8$ is not satisfied.

\section{Timed I/O Game}
\label{sec:game}

Our specification theory can be understood from a game theoretical point of view. It is an input-output game between a \emph{component} and an \emph{environment} that uses a \emph{coin} to break ties. The specification of a component (in the form of a TIOA or TIOTS) is built to encode the set of strategies possible for the component in the game (just like an NFA encodes a set of words). %We first develop a generic game framework on general TIOTSs, after which we restrict our framework to realisable TIOTSs.
%Generically the game can have $n$ players and an $n$-dice. But for the purpose of simplicity we restrict us to consider only binary games, i.e.

\vspace{-2mm}
\begin{itemize}
\item  Given two TIOTSs $\P$ and $\Q$ with identical alphabets, we say $\P$ is a partial unfolding~\cite{wang12} of $\Q$ if there exists a function $f$ from $S_{\P}$ to $S_{\Q}$ s.t. 1) $f$ maps $\top$ to $\top$, $\bot$ to $\bot$, and plain states to plain states, 2) $f(s^0_{\P})= s^0_{\Q}$, and 3) $p \ar{\alpha}_{\P} s \implies f(p) \ar{\alpha}_{\Q} f(s)$.
\item We say an acyclic TIOTS is a \emph{tree} if 1) there does not exist a pair of transitions in the form of $p \ar{a} p''$ and $p' \ar{d} p''$, 2) $p \ar{a} p'' \wedge p' \ar{b} p''$ implies $p=p'$ and $a=b$ and 3) $p \ar{d} p'' \wedge p' \ar{d} p''$ implies  $p=p'$. %$p \ar{\alpha} p'' \wedge p' \ar{\beta} p'' \wedge p \neq p'$ implies $\alpha, \beta \in \mathbb{R}^{>0} \wedge \alpha \neq \beta$.
\item We say an acyclic TIOTS is a \emph{simple path} if 1) $p \ar{a} s' \wedge p \ar{\alpha} s''$ implies $s'=s''$ and $a=\alpha$ and 2) $p \ar{d} s' \wedge p \ar{d} s''$ implies  $s'=s''$.
\item We say a simple path $\R$ is a \emph{run} of $\P$ if $\R$ is a partial unfolding of $\P$.
\end{itemize}
\vspace{-2mm}

\paragraph{Strategies.} A \emph{strategy} $\G$ is a deterministic tree TIOTS s.t. each plain state in $\G$ is ready to accept all possible inputs by the environment, but allows a single move (delay or output) by the component, i.e. $eb_{\G}(p) = I \uplus mv_{\G}(p)$ s.t. $mv_{\G}(p) = \{a\}$ for some $a \in O$ or $\{\} \subset mv_{\G}(p) \subseteq \mathbb{R}^{>0}$, where $eb_{\G}(p)$ denotes the set of enabled timed actions in state $p$ of LTS $\G$, and $mv_{\G}(p)$ denotes the unique component move allowed by $\G$ at $p$.

A TIOTS $\P$ \emph{contains} a strategy $\G$ if $\G$ is a partial unfolding of $(\P^{\bot})^{\top}$. The set of strategies\footnote{In this paper we use a set of strategies (say $\Pi$) to mean a set of strategies with identical alphabets} contained in $\P$ is denoted $stg(\P)$. Since it makes little sense to distinguish strategies that are isomorphic, we will freely use strategies to refer to their isomorphism classes and write $\G=\G'$ to mean $\G$ and $\G'$ are isomorphic.

\begin{figure}[t]
\begin{center}
\includegraphics[width=0.8\textwidth]{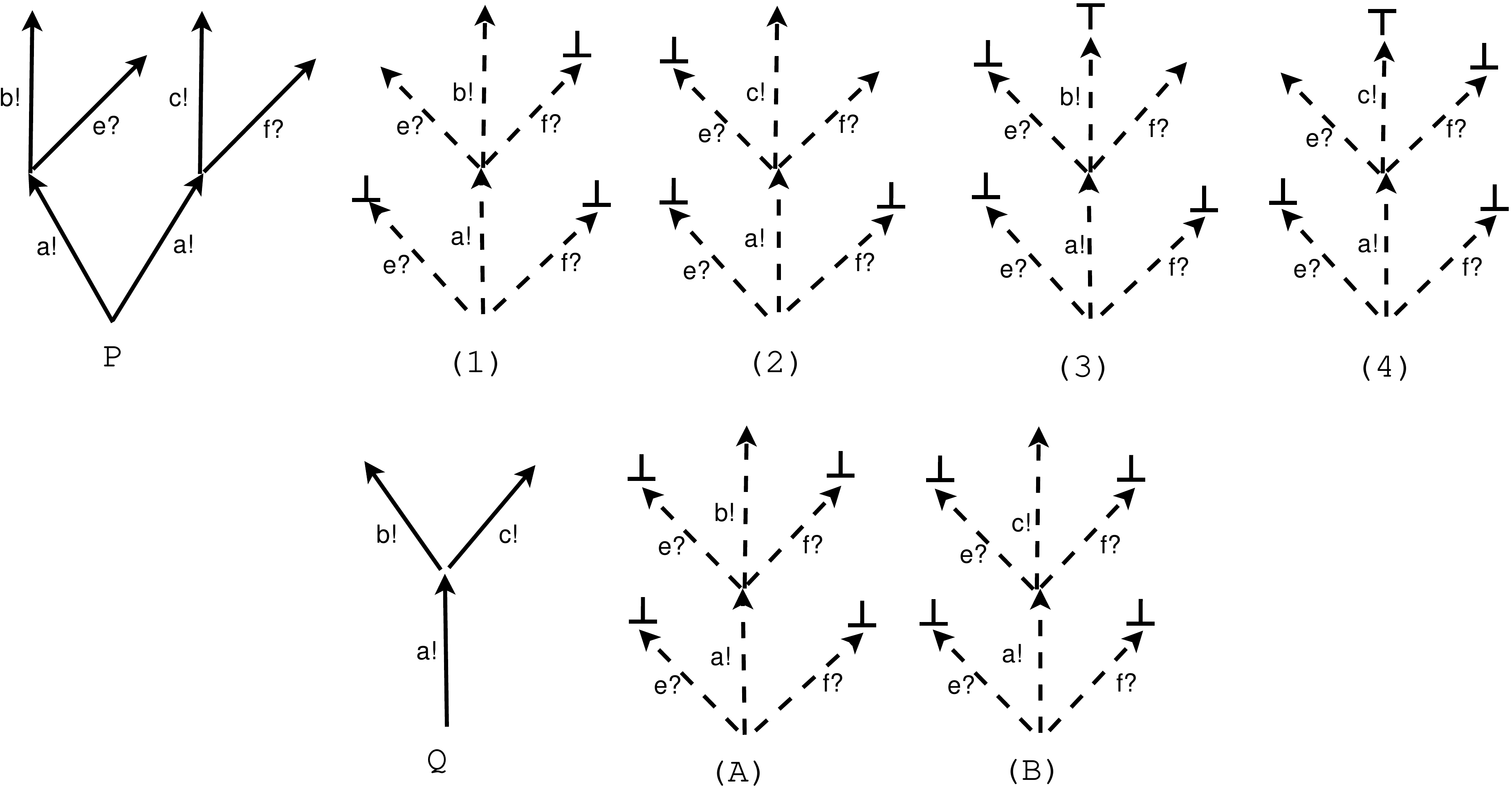}
\end{center}
\vspace{-5mm}
\caption{Strategy example.}
\vspace{-5mm}
\label{fig:strategy-equivalence-strong}
\end{figure}

Let us give some examples in Figure~\ref{fig:strategy-equivalence-strong}. For the sake of simplicity we use two untimed transition systems $\P$ and $\Q$, which have identical alphabets $I=\{e,f\}$ and $O=\{a,b,c\}$, to illustrate the idea of strategies. The transition systems use solid lines while strategies use dotted lines. Plain states are unmarked while the $\top$ and $\bot$ states are marked by $\top$ and $\bot$ resp.\footnote{To simplify drawing, multiple copies of $\top$ and $\bot$ are allowed but the self-loops on them are omitted.} We show four strategies of $\P$ and two strategies of $\Q$ on the right hand side of $\P$ and $\Q$ resp. in Figure~\ref{fig:strategy-equivalence-strong}. (They are not the complete sets of strategies for $\P$ and $\Q$.) Note that the strategies $3$ and $4$ own their existence to the $\top$ completion.

\paragraph{Comparing strategies.} When the game is played, the component tries to avoid reaching $\top$ while the environment tries to avoid reaching $\bot$. Different strategies in $stg(\P)$ vary in their effectiveness to achieve the objective. Such effectiveness can be compared if two strategies closely resemble each other: we say $\G$ and $\G'$ are \emph{affine} if $s^0_{\G} \dar{tt} p$ and $s^0_{\G'} \dar{tt} p'$ implies $mv_{\G}(p) = mv_{\G'}(p')$. Intuitively, it means $\G$ and $\G'$ propose the same move at the `same' states. For instance, the strategies $1$, $3$ and $A$ in Figure~\ref{fig:strategy-equivalence-strong} are pairwise affine and so are the strategies $2$, $4$ and $B$.

Given two affine strategies $\G$ and $\G'$, we say $\G$ is \emph{more aggressive} than $\G'$, denoted $\G \preceq \G'$, if 1) $s^0_{\G'} \dar{tt} \bot$ implies there is a prefix $tt_0$ of $tt$ s.t. $s^0_{\G} \dar{tt_0} \bot$ and 2) $s^0_{\G} \dar{tt} \top$ implies there is a prefix $tt_0$ of $tt$ s.t. $s^0_{\G'} \dar{tt_0} \top$. Intuitively, it means $\G$ can reach $\bot$ faster but $\top$ slower than $\G'$.  $\preceq$ forms a partial order over $stg(\P)$, or more generally, over any set of strategies with identical alphabets. For instance, strategy $A$ is more aggressive than $1$ and $3$, while strategy $B$ is more aggressive than $2$ and $4$.

% It is easy to see $\G \preceq \G'$ iff $G$ can be obtained from $G'$ by multiple applications of 1) collapsing sub-trees of $G'$ to $\bot$ and 2) expanding $\top$ in $G'$ into sub-trees.

When the game is played, the component $\P$ prefers to use the maximally aggressive strategies in $stg(\P)$\footnote{This is because our semantics is designed to preserve $\bot$ rather than $\top$.}. Thus two components that differ only in non-maximally aggressive strategies should be equated. We define the \emph{strategy semantics} of component $\P$ to be $[\P]_s = \{\G' \, | \, \exists \G\in stg(\P) : \G\preceq \G' \}$, i.e. the upward-closure of $stg(\P)$ w.r.t. $\preceq$.

%and 2) to collaborate with the coin to alure the environment into $\bot$.

%A \emph{strategy} is a realisable pre-strategy. The set of pre-strategies of a TIOTS $\P$ is denoted $stg(\P)$.
% via $f$ and 2) $f(p)=p' \wedge eb_{\P}(p') \cap tO \neq \{\}$ implies $eb_{\G}(p) \cap tO \neq \{\}$.

% and $\M{\P}$ to denote the magic-completion of an error-completed $\P$

%Pre-strategy: a deterministic tree-like partial unfolding s.t. in each state $x$ the possible move of the component is uniquely determined: i.e. it can either delay $eb(x) = I \uplus X$ where $X \subseteq \mathcal{R}^{>0}$ or $eb(x) = I \uplus \{a\}$ for some $a \in O$ is a singleton action set.

%Strategy is maximal pre-strategy free of timestop.

%\footnote{$(0..d]$ stands for the time interval from $0$ to $d$ excluding $0$ while $(0..d')$ stands for the time interval from $0$ to $d'$ excluding $0$ and $d'$.}
%those components which have \emph{complementary alphabets}. That is $I_{\P} = O_{\Q}$ and $O_{\P} = I_{\Q}$. $\Q$ encodes the set of strategies for the environment.

\vspace{-2mm}
\paragraph{Game rules.} When a component strategy $\G$ is played against an environment strategy $\G'$, at each game state (i.e. a product state $p_{\G} \times p_{\G'}$) $\G$ and $\G'$ each propose a move (i.e. $mv_{\G}(p_{\G})$ and $mv_{\G'}(p_{\G'})$). If one of them is a delay and the other is an action, the action will prevail. If both propose delay moves (i.e. $mv_{\G}(p_{\G}), mv_{\G'}(p_{\G'}) \subseteq \mathbb{R}^{\geq 0}$), the smaller one (w.r.t. set containment) will prevail.

Since a delay move proposed at a strategy state is the maximal set of possible delays enabled at that state, the next move proposed at the new state after firing the set must be an action move (due to time additivity). Thus a play cannot have two consecutive delay moves.

If, however, both propose action moves, there will be a tie, which will be resolved by tossing the coin. For uniformity's sake, the coin can be treated as a special component. A strategy of the coin is a function $h$ from $tA^*$ to $\{0,1\}$. We denote the set of all possible coin strategies as $H$.

%The definition of $\parallel$ over states (i.e. the \emph{state composition operator}) is in the first table above.
%\footnote{The formal definition of parallel composition on TIOTSs is in the next section.}

A play of the game can be formalised as a composition of three strategies, one each from the component, environment and coin, denoted $\G_{\P} \parallel_h \G_{\Q}$. At a current game state $p_{\P} \times p_{\Q}$, if the prevailing action is $\alpha$ and we have $p_{\P} \ar{\alpha} s'_{\P}$ and $p_{\Q} \ar{\alpha} s'_{\Q}$, then the next game state is  $s_{\P} \parallel s_{\Q}$. The play will stop when it reaches either $\top$ or $\bot$. The composition will produce a simple path $\R$ that is a run of $\P \parallel \Q$. Since $\P \parallel \Q$ gives rise to a \emph{closed system} (i.e. the input alphabet is empty), a run of $\P \parallel \Q$ is a strategy of $\P \parallel \Q$.

This is crucial since it reveals that strategy composition of $\P$ and $\Q$ is closely related to their parallel composition:  $stg(\P \parallel \Q) = \{\G_{\P} \parallel_h \G_{\Q} \, | \, \G_{\P} \in stg(\P), \G_{\Q} \in stg(\Q) $ and $h \in H\}$.

%Such play of game also give us a new perspective to interpret parallel composition: Note that $\P \parallel \Q$ gives us a closed component

\paragraph{Parallel composition.} Strategy composition, like component (parallel) composition, can be generalised to any pair of components $\P$ and $\Q$ with \emph{composable alphabets}. That is, $O_{\P} \cap O_{\Q} = \{\}$. For such $\P$ and $\Q$, $\G_{\P} \parallel_h \G_{\Q}$ gives rise to a tree rather than simple path TIOTS. That is, at each game state $p_{\P} \times p_{\Q}$, besides firing the prevailing $\alpha \in tO_{\P} \cup tO_{\Q}$, we need also to fire 1) all the synchronised inputs, i.e. $e \in I_{\P} \cap I_{\Q}$, and reach the new game state $s_{\P} \parallel s_{\Q}$ (assuming $p_{\P} \ar{e} s_{\P}$ and $p_{\Q} \ar{e} s_{\Q}$) and 2) all the independent inputs, i.e. $e \in (I_{\P} \cup I_{\Q}) \setminus (A_{\P} \cap A_{\Q})$, and reach the new game state $s_{\P} \times p_{\Q}$ or $p_{\P} \times s_{\Q}$. It is easy to verify that $\G_{\P} \parallel_h \G_{\Q}$ is a strategy of $\P \parallel \Q$.

\paragraph{Conjunction/disjunction.} Besides strategy composition, \emph{strategy conjunction} ($\&$) and \emph{strategy disjunction} ($+$) are also definable. They are binary operators defined only on pairs of affine strategies. We define $\G \& \G' = \G \wedge \G'$ and $\G + \G' = \G \vee \G'$. Note that if $\G$ and $\G'$ are not affine, $\G \wedge \G'$ and $\G \vee \G'$ do not necessarily produce a strategy. For instance the disjunction of the strategies $1$ and $2$ in Figure~\ref{fig:strategy-equivalence-strong} will produce a transition system that stops to output after the $a$ transition.

\paragraph{Refinement.} Strategy semantics induce an equivalence on TIOTSs. That is, $\P$ and $\Q$ are \emph{strategy equivalent} iff $[\P]_s = [\Q]_s$. However, strategy equivalence is too fine for the purpose of \emph{substitutive refinement} (cf Definition~\ref{defn:op-refine}). For instance, transition systems $\P$ and $\Q$ in Figure~\ref{fig:strategy-equivalence-strong} are substitutively equivalent, but are not strategy equivalent, because $1$, $2$, $3$ and $4$ are strategies of $\Q$ (due to upward-closure w.r.t. $\preceq$), but $A$ and $B$ are not strategies of $\P$.

However, we demonstrate that \emph{substitutive equivalence is reducible to strategy equivalence} providing we perform \emph{disjunction closure} on strategies. %followed by \emph{error-backpropagation closure} on strategies., and 2) keep only the maximally aggressive strategies.

\begin{lemma}
Given a pair of affine component strategies $\G_0$ and $\G_1$, $\G_0 \parallel_h \G$ and $\G_1 \parallel_h \G$ are $\bot$-free for some environment strategy $\G$ and $h \in H$ iff $\G_0 + \G_1 \parallel_h \G$ is $\bot$-free.
\end{lemma}

%\begin{lemma}
%Given a pair of affine (component) strategies $\G_0$ and $\G_1$ and an environment strategy $\G$, $\G_0 \parallel_h \G$ and $\G_1 \parallel_h \G$ are $\bot$-free for all $h \in H$ iff $\G_0 + \G_1 \parallel_h \G$ is $\bot$-free for all $h \in H$.
%\end{lemma}

We say $\Pi^+$ is a \emph{disjunction closure} of $\Pi$ iff it is the least superset of $\Pi$ s.t. $\G + \G' \in \Pi^+$ for all pairs of affine strategies $\G, \G' \in \Pi^+$. It is easy to see the disjunction closure operation preserves the upward-closedness of strategy sets.

\begin{theorem}
Given TIOTSs $\P$ and $\Q$,  $\P \sqsubseteq \Q$ iff $[\Q]_s^+ \subseteq [\P]_s^+$.
\end{theorem}

For instance, the disjunction of strategies $1$ and $3$ produces $A$, while the disjunction of strategies $2$ and $4$ produces $B$. Thus $[\P]_s^+ = [\Q]_s^+$,

\paragraph{Relating operational composition to strategies.} The operations of parallel composition, conjunction and disjunction defined on the operational models of TIOTSs (Section~\ref{sec:optheory}) can be characterised by simple operations on strategies in the game-based setting.

\begin{lemma}
For $\parallel$-composable TIOTSs $\P$ and $\Q$, $[\P \parallel \Q]_s^+ = \{ \G_{\P \parallel \Q} \, | \, \exists \G_{\P} \in [\P]_s^+, \G_{\Q} \in [\Q]_s^+, h \in H : \G_{\P} \parallel_h \G_{\Q} \preceq \G_{\P \parallel \Q} \}$.
\end{lemma}
%\begin{proof}
%Follow from 1) state $\parallel$ is distributive over state $\vee$ and 2) $stg(\P \parallel \Q) = \{\G_{\P} \parallel_h \G_{\Q} \, | \, \G_{\P} \in stg(\P), \G_{\Q} \in stg(\Q) $ and $h \in H\}$.
%\end{proof}

\begin{lemma}
For $\vee$-composable TIOTSs $\P$ and $\Q$, $[\P \vee \Q]_s^+= ([\P]_s^+ \cup [\Q]_s^+)^+$.
\end{lemma}

\begin{lemma}
For $\wedge$-composable TIOTSs $\P$ and $\Q$, $[\P \wedge \Q]_s^+= [\P]_s^+ \cap [\Q]_s^+$.
\end{lemma}

%\G_{\P} \in [\P]_s^+, \,

\begin{lemma}
For $\%$-composable TIOTSs $\P$ and $\Q$, $[\P \% \Q]_s^+= \{\G_{\P \% \Q}  \, | \, \forall \G_{\Q} \in [\Q]_s^+, h \in H: \G_{\P \% \Q} \parallel_h \G_{\Q} \in  [\P]_s^+ \}$.
\end{lemma}

Thus conjunction and disjunction are the join and meet operations and quotient produces the coarsest TIOTS s.t. $(\P_0 \% \P_1) \para \P_1$ is a refinement of $\P_0$.

\begin{lemma}
For any TIOTS $\P$, $[\P^{\neg}]_s^+= \{\G_{\P^{\neg}}  \, | \, \forall \G_{\P} \in [\P]_s^+, h \in H: \G_{\P^{\neg}} \parallel_h \G_{\P} $\ is $\bot$-free$ \}$.
\end{lemma}

\begin{theorem}
$\simeq$ is a congruence w.r.t. $\parallel$, $\vee$, $\wedge$ and $\%$ subject to composability.
\end{theorem}

\paragraph{Summary.} Strategy semantics has given us %two weakest $\bot$-preserving congruences (i.e. $[\P]_s^{\vee}$ and $[\P]_n$)
a weakest $\bot$-preserving congruence (i.e. $[\P]_s^+$) for timed specification theories based on operators for (parallel) composition, conjunction, disjunction and quotient. Strategy semantics captures nicely the game-theoretical nature as well as the operational intuition of the specification theories. However, in a more declarative manner, the equivalence can also be characterised by timed traces, as we see in the next section.

\section{Declarative Specification Theory}
\label{sec:ts}

In this section, we develop a compositional specification theory based on timed traces. We introduce the concept of a timed-trace structure, which is an abstract representation for a timed component. The timed-trace structure contains essential information about the component, for checking whether it can be substituted with another in a safety and liveness preserving manner.

%\paragraph{Notation.} We write $\mathbb{R}$ for the set of non-negative real numbers, including $0$.

%\begin{definition}[Timed-trace]
%\label{defn:ts-tt}
%For a set of actions $\A$, a timed-trace over $\A$ is a non-empty alternating sequence of time-values and actions commencing with a time-value. A timed-trace is thus an element of the set
%$\mathbb{R}(\A\ \mathbb{R})^{*}(\e + \A)$, which we denote by $\TT{\A}$.
%\end{definition}

%Note that a timed-trace does not stipulate that the time-values are cumulative. This means that a time-value corresponds to the time sojourned between adjacent actions. In this section, we do not stipulate that a timed-system must respect properties such as progress of time and non-Zenoness. Thus we concern ourselves with arbitrary timed-traces.

%\paragraph{Notation.} We let $a$ range over actions, while $x$ ranges over $\mathbb{R}$.

Given any TIOTS $\P= \langle I, O, S, s^0, \rightarrow \rangle$, we can extract three sets of traces from $(\P^{\bot})^{\top}$: $TP$ (plain traces) is a set of timed traces leading to plain states, $TE$ (error traces) a set of timed traces leading to $\bot$ and $TM$ (magic traces) a set of timed traces leading to $\top$. The three sets contain sufficient but not necessary information for our substitutive refinement, which is designed to preserve $\bot$ rather than $\top$. For instance, adding any trace $tt \in TE$ to $TP$ should not change the semantics of the component; similarly it is true for removing any trace $tt \in TP$ from $TM$. Based on a slight abstraction of the three sets we can define a \emph{triple-trace structure} as the semantics of $\P$.

\begin{definition}[Triple-trace structure]
$\TT{\P} := (I, O, TT, TR, TE)$, where $TT := TE \cup TP \cup TM$ is the set of all traces and $TR := TE \cup TP$ the set of realisable traces.
\end{definition}

Obviously, $TE$ is extension-closed. $TT$ is non-empty and prefix-closed. $TR$ is prefix-closed and \emph{fully branching}\footnote{This is due to $\top/\bot$-completion.} w.r.t. $TT$ (i.e. $tt \cat \langle \alpha \rangle \in TT$ for all $tt \in TR$ and $\alpha \in tA$). $TT \setminus TR$ is time-extension closed (i.e. $tt \in X \implies tt \cat \langle d\rangle \in X$) and any pair of traces from $TT \setminus TR$ that are related by extension are related by time-extension.

%our congruences cannot tell it from the case that $tt$ only belongs to $TE$ (and similarly for $TP$ and $TM$ to $TP$).

%where $TP$ is a set of timed traces leading to plain states, $TE$ a set of timed traces leading to $\bot$ and $TM$ a set of timed traces leading to $\top$\footnote{Note that $\top$ and $\bot$ are both chaotics states.}. Their union $TT := TE \cup TP \cup TM$ is the set of all timed traces while $TS := TE \cup TP$ the set of non-magical traces.
% and $last(tt) \in tO$ for all $tt \in TT \setminus TR$

From hereon let $\P_0$ and $\P_1$ be two TIOTSs with triple trace structures $\TT{\P_i} := (I_i, O_i, TT_i, TR_i, TE_i)$ for $i \in \{0,1\}$. Define $\bar{i}=1-i$. %If $TE_0$ is empty we say that $\P_0$ is \emph{free of assumptions} (on the environment). Note that assumption-freeness on triple-trace structures is very different from the notion of $\bot$-freeness on TIOTSs.

%A triple-trace structure is \emph{weakly realisable} iff $TR$ is \emph{receptive}. A timed trace set $TR$ is \emph{receptive} iff, given any $tt \in TR$, $tt \cat \langle e \rangle \in TR$ for all $e \in I$ and there exists $tt \cat w \in TR$ for all $d \in R^{>0}$ such that $w \in tO^* \land l(w) = d$. Thus we can derive $TT$ from $TR$ by a $\top$-completion like operation, i.e. $TT=TR^{\top}$.

%$TP$ is prefix-closed. $TR$ must be non-empty, \emph{fully branching} and prefix-closed. A timed trace set $TT$ is \emph{fully branching} iff,

The substitutive refinement relation $\sqsubseteq$ in Section~\ref{sec:optheory} can equally be characterised by means of trace containment. Consequently, $\TT{\P_0}$ can be regarded as providing an alternative encoding of the set $[\P_0]_s^+$ of strategies.

\begin{theorem}
%Given TIOTSs $\P_0$ and $\P_1$ s.t. $\TT{\P_i} := (I, O, TT_i, TR_i, TE_i)$ for $i \in \{0,1\}$,
$\P_0 \sqsubseteq \P_1$ iff $TT_1 \subseteq TT_0$, $TR_1 \subseteq TR_0$ and $TE_1 \subseteq TE_0$.
\end{theorem}

We are now ready to define the timed-trace structure semantics for the operators of our specification theory. Intuitively, the timed-trace semantics mimic the synchronised product of the operational definitions in Section~\ref{sec:optheory}. An important fact utilised in formulating these operations on traces is that for any trace $tt \in tA^*$ and TIOTS $\P$, either $tt$ is a trace of $\P$ or there is some prefix $tt_0$ of $tt$ s.t. $tt_0$ is an error or magic trace of $\P$.

\paragraph{Parallel composition.} The idea behind parallel composition is that the projection of any trace in the composition onto the alphabet of one of the components should be a trace of that component.

\begin{proposition}
If $\P_0$ and $\P_1$ are $\parallel$-composable, then $\TT{\P_0 \parallel \P_1} = (I, O, TT, $ $TR, TE)$ where $I=(I_0\cup I_1) \setminus O$, $O=O_0\cup O_1$ and the trace sets are given by:
\begin{itemize}
\item $TE = \{ tt | tt \upharpoonright tA_i \in TE_i \wedge tt \upharpoonright tA_{\bar{i}} \in TR_{\bar{i}} \}  \cdot tA^*$
\item $TR = TE \uplus \{ tt | tt \upharpoonright tA_i \in (TR_i \setminus TE_i) \wedge tt \upharpoonright tA_{\bar{i}} \in (TR_{\bar{i}} \setminus TE_{\bar{i}}) \} $
\item $TT = TR \uplus \{ tt | tt \upharpoonright tA_i \in (TT_i \setminus TR_i)  \wedge tt_0  < tt \upharpoonright tA_{\bar{i}} \implies tt_0  \in (TR_{\bar{i}} \setminus TE_{\bar{i}}) \} \cdot  \mathbb{R}^{\geq 0}$.
\end{itemize}
\end{proposition}

The above says $tt$ is an error trace if the projection of $tt$ on one component is an error trace while the projection of $tt$ on the other component is not a magic trace. $tt$ is a realisable trace if $tt$ is either an error trace or a plain trace. $tt$ is a plain trace if the projection of $tt$ on both components are plain traces. Finally, $tt$ is a magic trace if its projection on one component is a magic trace, while the projection of all strict prefixes of $tt$ on the other component is a plain trace.

%Given two composable specifications $\P_0$ and $\P_1$, $\BT{\P_0 \parallel \P_1}$ definition follows from that of  $\DT{\P_0 \parallel \P_1}$ since $\parallel$ coincides with $\parallel_r$ and preserves realisability.

% with $\DT{\P_0 \parallel \P_1} = (I, O, TR, TE)$, we define $\BT{\P_0 \parallel \P_1} := (I, O, TR \cup TE^{ebp}, TE^{ebp})$.

%$\BT{\P_i} := (I_i, O_i, TR^{ebp}_i, TE^{ebp}_i)$ for $i \in \{0,1\}$, we define $\BT{\P_0 \parallel \P_1} = (I, O, TR^{ebp}, TE^{ebp})$, where

%\begin{itemize}
%\item $TE^{ebp} = \{ tt | tt \upharpoonright tA_i \in TE^{ebp}_i \wedge tt \upharpoonright tA_{\bar{i}} \in TR^{ebp}_{\bar{i}} \}  \cdot tA^*$
%\item $TR^{ebp} = TE^{ebp} \uplus \{ tt | tt \upharpoonright tA_i \in (TR^{ebp}_i \setminus TE^{ebp}_i) \wedge tt \upharpoonright tA_{\bar{i}} \in (TR^{ebp}_{\bar{i}} \setminus TE^{ebp}_{\bar{i}}) \} $
%\end{itemize}

\paragraph{Disjunction.} From any composite state in the disjunction of two components, the composition should only be willing to accept inputs that are accepted by both components, but should accept the union of outputs. After witnessing an output enabled by only one of the components, the disjunction should behave like that component. Because of the way that $\bot$ and $\top$ work in Table~\ref{table:composition}, this loosely corresponds to taking the union of the traces from the respective components.

\begin{proposition}
If $\P_0$ and $\P_1$ are $\vee$-composable, then $\TT{\P_0 \vee \P_1} = (I, O, TR_0 \cup TR_1 \cup TM,  TR_0 \cup TR_1, TE_0 \cup TE_1)$, where $I = I_0 = I_1$, $O=O_0 = O_1$ and $TM = \{ tt | tt \in (TT_i \setminus TR_i)  \wedge \exists tt_0  \leq tt : tt_0  \in (TT_{\bar{i}} \setminus TR_{\bar{i}}) \} \cdot  \mathbb{R}^{\geq 0}$.
\end{proposition}

Essentially, $tt$ is a magic trace if it is a magic trace on one component while one of its prefixes is a magic trace on the other component. The realisable and error traces are simply the union of the corresponding traces on $\P_0$ and $\P_1$.

%, where $TE = TE_0 \cup TE_1$, %\{ tt | tt \in TE_i \wedge \exists tt_0 \leq tt :  tt_0 \in TT_{\bar{i}} \cup (TT_{\bar{i}})  \}  \cdot tA^*$
%$TR = TR_0 \cup TR_1$ %  \uplus \{ tt | tt \upharpoonright tA_i \in (TR_i \setminus TE_i) \wedge tt \upharpoonright tA_{\bar{i}} \in (TR_{\bar{i}} \setminus TE_{\bar{i}}) \} $
%and $TT = TT_0 \cup TT_1$. %TR \uplus \{ tt | tt \upharpoonright tA_i \in (TT_i \setminus TR_i)  \wedge tt_0  < tt \upharpoonright tA_{\bar{i}} \implies tt_0  \in (TR_{\bar{i}} \setminus TE_{\bar{i}}) \} \cdot  \mathbb{R}^{\geq 0}$
%\end{itemize}

%Given two specifications with identical alphabets $\P_0$ and $\P_1$, $\BT{\P_0 \vee \P_1}$ definition follows from that of $\DT{\P_0 \vee \P_1}$ since $\vee$ coincides with $\vee_r$ and preserves realisability.
% s.t. $\BT{\P_i} := (I_i, O_i, TR_i, TE_i)$ for $i \in \{0,1\}$, we define $\BT{\P_0 \parallel \P_1} = (I, O, TR \cup TE^{ebp}, TE^{ebp})$, where $TE = TE_0 \cup TE_1$ and $TR = TR_0 \cup TR_1$.

\paragraph{Conjunction.} Similarly to disjunction, from any composite state in the conjunction of two components, the composition should only be willing to accept outputs that are accepted by both components, and should accept the union of inputs, until a stage when one of the component's input assumptions has been violated, after which it should behave like the other component. Because of the way that both $\bot$ and $\top$ work in Table~\ref{table:composition}, this essentially corresponds to taking the intersection of the traces from the respective components.

\begin{proposition}
If $\P_0$ and $\P_1$ are $\wedge$-composable, then $\TT{\P_0 \wedge \P_1} = (I, O, (TR_0 \cap TR_1) \cup TM,$  $TR_0 \cap TR_1, TE_0 \cap TE_1)$, where $I = I_0 = I_1$, $O=O_0 = O_1$ and $TM = \{ tt | tt \in (TT_i \setminus TR_i)  \wedge tt_0  < tt \implies tt_0  \in TR_{\bar{i}} \} \cdot  \mathbb{R}^{\geq 0}$.
\end{proposition}

A trace $tt$ is a magic trace if it is a magic trace on one of the components, and all strict prefixes of the trace are realisable by the other component. The realisable and error traces are simply the intersection of the corresponding traces on $\P_0$ and $\P_1$.

\paragraph{Quotient.} Quotient ensures its composition with the second component is a refinement of the first. Given the synchronised running of $\P_0$ and $\P_1$, if $\P_0$ is in a more refined state than $\P_1$, the quotient will try to rescue the refinement by taking $\top$ as its state (so that its composition with $\P_1$'s state will refine $\P_0$'s). If $\P_0$ is in a less or equally refined state than $\P_1$'s, the quotient will take the worst possible state without breaking the refinement.
\label{sec:ts-quotient}

\begin{proposition}
If $\P_0$ dominates $\P_1$, then $\TT{\P_0 \% \P_1} = (I, O, TT, TR, TE)$, where $I= I_0 \cup O_1$, $O=O_0\setminus O_1$, and the trace sets satisfy:

\begin{itemize}
\item $TE = \{ tt | (tt \in TE_0 \wedge tt_0 < tt \implies tt_0 \upharpoonright tA_1 \notin TE_1) \vee
                    (tt \upharpoonright tA_1 \in (TT_1 \setminus TR_1) \wedge tt_0 < tt \implies tt_0 \notin TT_0 \setminus TR_0) \}  \cdot tA^*$
\item $TR = TE \uplus \{ tt | tt \in (TR_0 \setminus TE_0) \wedge tt \upharpoonright tA_1 \in (TR_1 \setminus TE_1) \} $
\item $TT = TR \uplus \{ tt | (tt \in (TT_0 \setminus TR_0) \wedge tt_0 \leq tt \implies tt_0 \upharpoonright tA_1 \in TR_1) \vee
                    (tt \upharpoonright tA_1 \in TE_1 \wedge tt_0 \leq tt \implies tt_0 \notin TE_0) \}$.
\end{itemize}
\end{proposition}

The above says $tt$ is an error trace if either 1) $tt$ is an error trace in $\P_0$, but the projection of any strict prefix of $tt$ on $\P_1$ is not an error trace, or 2) the projection of $tt$ on $\P_1$ is a magic trace, but no strict prefix of $tt$ is a magic trace in $\P_0$. $tt$ is a magic trace if either 1) $tt$ is a magic trace in $\P_0$, but the projection of any prefix of $tt$ is not a magic trace in $\P_1$, or 2) the projection of $tt$ on $\P_1$ is an error trace, but no prefix of $tt$ is an error trace in $\P_0$.

%Given two specifications $\P_0$ and $\P_1$ s.t. the alphabets of $P_0$ \emph{dominate} those of $P_1$, $\P_0 \%_r \P_1$ is defined and $\TT{\P_0 \%_r \P_1} = (I, O, TT, TR, TE)$, we define $\BT{\P_0 \%_r \P_1} = (I, O, TR \setminus TM^{mbp}, TE \setminus TM^{mbp})$.

Mirroring of triple trace structures is straightforward: $\TT{\P_0}^\neg = (O_0, I_0, $ $TT_0,$ $ TT_0 \setminus TE_0, TT_0 \setminus TR_0)$. This is because dealing with traces means we have implicit determinism, so we can skip the determinisation step. Consequently, quotient can also be defined as the derived operator $(\TT{\P_0}^\neg \parallel \TT{\P_1})^\neg$.

%\paragraph{Congruences.} It can be shown that substitutive equivalence of the operations in the specification theory forms a congruence.

\section{Comparison with Related Works}
\label{sec:comp}

Based on linear-time, our timed theory owes much to the pioneering work of trace theories in asynchronous circuit verification, such as Dill's trace theory \cite{dill-trace-theory}.
%Our theory can be reformulated for the operators of parallel, hiding and renaming as a timed-extension of Dill's trace theory (prefix-closed trace structure)\footnote{According to our best knowledge it will be the first such timed extension.}.
Our mirror operator is essentially a timed extension of the mirror operator from asynchronous circuit verification. The definition of quotient based on mirroring (for the untimed case) was first presented by Verhoeff as his Factorisation Theorem~\cite{verhoeff-thesis}.

%Our work is also deeply influenced by the work of [Alfaro]. We use the timed-game work of [Alfaro] with some adaptations. Firstly, rather than being treated as a timed-game graph, a TIOTS or component specification is regarded as a set of component strategies. We adopt most of the game rules in [Alfaro], but due to our formulation of proposed delay moves as maximal delay allowed by a strategy, a play of our game cannot have consecutive delay moves.\marginpar{Chris: I don't think we insist on alternation between actions and delays?} This enables us to avoid the complexity of time-blocking strategies and blame assignment\footnote{Note that this does not exclude zeno behaviours (infinite action moves within finite time) in our play, which are not regarded as abnormal behaviours in our semantics.}. \marginpar{Chris: but it doesn't avoid Zenoness} Secondly, we do not use timestop (or semi-timestop) to model time errors (i.e. bounded-liveness errors). Rather, we introduce the explicit inconsistent state $\bot$ to model both time and immediate (i.e. safety) errors. Timestop is used to model the magic state, which can simplify the definition of operators like conjunction and quotient. Conjunctions and quotient will introduce new timestop and semi-timestop. But it is not treated as errors. Instead it just introduces some unrealistic strategies which can be safely removed. This second point enables us to avoid the complexity of two transition relation and well-formedness of timed interfaces.

Our work is also deeply influenced by the work of~\cite{henzinger-timedia} on timed games, with some modifications.
%work of~\cite{henzinger-timedia} with some adaptations.
Firstly, a TIOTS is regarded as a set of component strategies, rather than a timed game graph. We adopt most of the game rules in~\cite{henzinger-timedia}, except that, due to our requirement that proposed delay moves are maximal delays allowed by a strategy, a play
%of our game
cannot have consecutive delay moves.
%\marginpar{Chris: I don't think we insist on alternation between actions and delays?}
This enables us to avoid the complexity of time-blocking strategies and blame assignment, but does not ensure non-Zenoness\footnote{Zeno behaviours (infinite action moves within finite time) in a play are not regarded as abnormal behaviours in our semantics.}.
%\marginpar{Chris: but it doesn't avoid Zenoness}
Secondly, we do not use timestop/semi-timestop to model time errors (i.e. bounded-liveness errors). Rather, we introduce the explicit inconsistent state $\bot$ to model both time and immediate (i.e. safety) errors. Timestop is used to model the magic state, which can simplify the definition of parallel, conjunction and quotient and enables us to avoid the complexity of having two transition relations and well-formedness of timed interfaces. %Conjunctions will introduce new timestop and semi-timestop, but instead of being treated as errors, these introduce some unrealistic strategies which can be safely removed. This second point enables us to avoid the complexity of having two transition relations and well-formedness of timed interfaces.

Last but not least, our work is related to~\cite{larsen-timedio}, as both devise a complete timed specification theory. The major differences lie in the use of timed alternating simulation as refinement in~\cite{larsen-timedio}, while ours is linear-time. An advantage of our work is that refinement is the weakest congruence preserving inconsistency, while beneficial in~\cite{larsen-timedio} is the algorithmic efficiency of branching-time simulation checking. Moreover,~\cite{larsen-timedio} has fully implemented the timed-game algorithms.

%Other related works that we do not consider here %, which will not be compared with,
We briefly mention other related works, which
include timed modal transition systems~\cite{Bert09,Cerans93}, the timed I/O model~\cite{Kaynar} and embedded systems~\cite{Thiele06,Lee07}.

\section{Conclusions}
\label{sec:concl}

We have formulated a rich compositional specification theory for components with real-time constraints based on a linear-time notion of substitutive refinement. The operators of hiding and renaming can also be defined, according to our past experiences~\cite{WangKwiat07}. We believe that our theory can be reformulated as a timed extension of Dill's trace theory \cite{dill-trace-theory}. Future work will include an investigation of realisability and assume-guarantee reasoning.

\paragraph{Acknowledgments.} The authors are supported by EU FP7 project CONNECT and ERC Advanced Grant VERIWARE.

%\begin{thebibliography}{10}\label{bibliography}
%\end{thebibliography}

\bibliographystyle{splncs}
\bibliography{timed-bib}

\appendix

\end{document}